\documentclass[usenatbib,usegraphicx,times]{mn2e}

\title[Why is the Moon synchronously rotating?]{Why is the Moon synchronously rotating?}
\author[Valeri V. Makarov]{Valeri V. Makarov$^{1}$\thanks{E-mail:
vvm@usno.navy.mil} 
\\
$^{1}$US Naval Observatory, 3450 Massachusetts Ave NW, Washington DC 20392-5420, USA}
\begin{document}

\date{Accepted . Received ; in original form }

\pagerange{\pageref{firstpage}--\pageref{lastpage}} \pubyear{2012}

\maketitle

\label{firstpage}

\begin{abstract}
If the Moon's spin evolved from faster prograde rates, it could have been captured into a higher
spin-orbit resonance than the current 1:1 resonance. At the current value of orbital eccentricity,
the probability of capture into the 3:2 resonance is as high as 0.6, but it strongly depends on the temperature and
average viscosity of the Moon's interior. A warmer, less viscous Moon on a higher-eccentricity orbit
is even more easily captured into supersynchronous resonances. 
We discuss two likely scenarios for the present spin-orbit state:
a cold Moon on a low-eccentricity orbit and a retrograde initial rotation.
\end{abstract}


\section{Introduction}
The origin of the Moon and the circumstances of its dynamical evolution remain unclear to date,
despite the numerous studies on this issue. The low eccentricity ($e=0.0549$) and the exactly 
synchronous rotation
suggest a nearly perfect equilibrium state, which is the end-point of dynamical spin-orbit evolution
\citep{hut}. They also point at a protracted history of dynamical interaction with the Earth and the Sun,
in which tidal dissipation undoubtedly played a crucial role. In the framework of the giant impact theory of
Moon's origin, tidal dissipation is responsible for the expansion of the orbit and damping of eccentricity
\citep{cuk}. Numerical simulations validating this hypothesis have been based on much 
simplified and {\it ad hoc}
models of tides, which should not be used for planets and moons of terrestrial composition \citep{em}.
In recent years, a more realistic model of tidal dissipation in solid bodies was proposed, which combines
the viscoelastic response with the inelastic creep \citep{el, ew, efr2}. In the framework of this model,
the capture of Mercury into the current 3:2 spin-orbit resonance becomes a likely and natural outcome even
without involving the core-mantle friction and episodes of high orbital eccentricity \citep{ma12}, which has been a difficult issue for the previous theories. For example, the constant time lag (CTL) model predicts capture probabilities into 3:2 of less than $0.1$ for
a wide range of parameters \citep{gold}.
Within the Efroimsky (2012) model, the secular tidal torque is rendered by a Darwin-Kaula expansion over the Fourier modes of the tide. Each term of the series assumes, in the vicinity of the appropriate resonance, the shape of a kink. As the tidal mode corresponding to the term transcends zero, the term swiftly, but continuously, changes its sign
vanishing at the resonance frequency. This behaviour makes the Efroimsky torque very efficient
at trapping the spin rate into resonances of higher order than the synchronous rotation. Assuming a terrestrial
composition for the super-earth GJ 581d, \citet{mabe} concluded that this potentially habitable exoplanet
is more likely to be found at a 2:1 spin-orbit resonance rather than 1:1. In this paper, I reassess the 
probabilities of capture of the Moon into supersynchronous resonances.

\section{Secular tidal torque}
\begin{figure*}
\begin{minipage}{82mm}
\includegraphics[width=82mm]{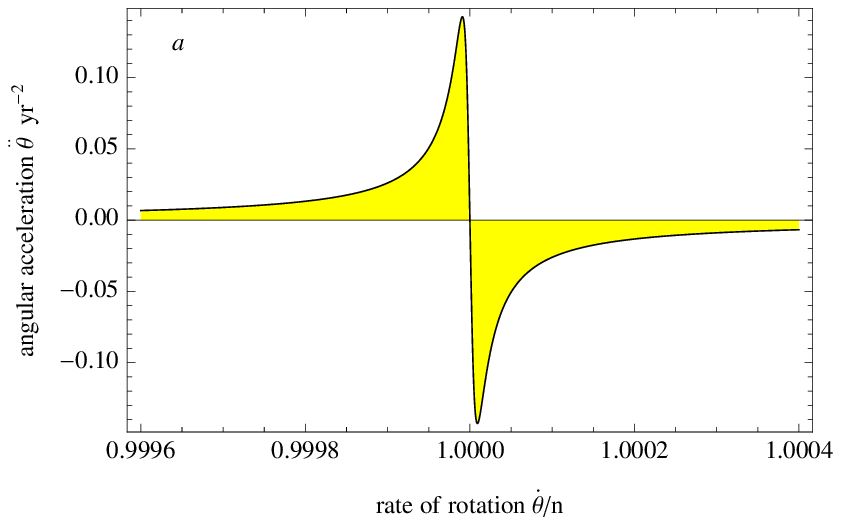}
\end{minipage}\hspace{2pc}
\begin{minipage}{82mm}
\includegraphics[width=82mm]{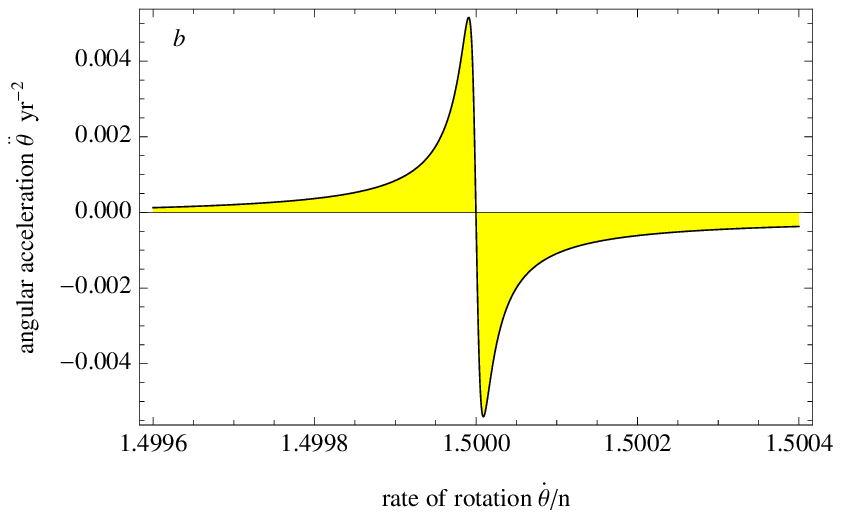}
\end{minipage}
\caption{Angular acceleration of the Moon caused by the secular component of the polar tidal torque in the
vicinity of a) 1:1 resonance; b) 3:2 resonance. }
\label{kink.fig}
\end{figure*}

Comprehensive equations for the polar tidal torque (i.e., the component directed along the axis of rotation),
including the fast oscillating terms, can be found, for example, in \citep{mabe}. They are not reproduced here
for brevity. The secular term of the torque, $K_c$, is strongly dependent on tidal frequency in the narrow
vicinity of spin-orbit resonances $(2+q)n=2\dot\theta$ for integer $q$, where $n$ is the mean motion and
$\dot\theta$ is the sidereal spin rate. 
The characteristic kink-shape of the near-resonant torque is present at both 1:1 and 3:2 resonances of the
Moon, but the former is by far larger than the latter and the other, higher-order resonances 
(Fig.~\ref{kink.fig}). The torque
is positive, or accelerating, at forcing frequencies below the resonance value and negative, or
decelerating, above it. The very steep decline
between the two peaks occupies a narrow band of frequencies for realistic rhelogies. Despite the relatively
small amplitude of the kink (compared to a typical amplitude of the triaxiality-caused torque), it acts
as an efficient trap for a planet trying to traverse the resonance. 

The numerical simulations presented in this paper were performed with physical parameters listed in Table 1.
The Andrade parameter $\alpha$ has been measured for a diverse list of materials, including
silicates, metals, and ices, and found to vary within a fairly narrow range of $0.14$ -- $0.3$.
The value $0.2$ estimated for hot silicate rocks is used in this paper. The unrelaxed rigidity
modulus $\mu$ takes values between $0.62$ and $0.68$ times $10^{11}$ Pa \citep{eck}. The assumed value
here is $0.65\cdot 10^{11}$ Pa. The most defining parameter in this model
is the Maxwell time $\tau_M$, which I varied in my analysis between 8 yr and 500 yr (approximately, the Earth's
value). The former value corresponds to a warmer satellite with less internal viscosity. As explained in
\citep{me}, the choice of a small Maxwell time for the Moon, only 8 years, may be justified by the likely 
presence of a high percentage of partial melt in the lower lunar mantle. The presence of partial melt 
indirectly follows 
from the modeling carried out by \citet{web} and also from an earlier study by \citet{nak}.

 \begin{table*}
 \centering
 \caption{Parameters of the tidal model.}
 \begin{tabular}{@{}lrrr@{}}
 \hline
            &                 &       &\\
   Name     &  Description    & Units & Values\\
 \hline
 $\xi$ & \dotfill moment of inertia coefficient &   &  2/5\\
 $R$   & \dotfill radius of planet              & m & $1.737\cdot 10^6$\\
 $M_2$ & \dotfill mass of the perturbed body (Moon)& kg & $7.3477\cdot 10^{22}$\\
 $M_1$ & \dotfill mass of the perturbing body (Earth)& kg & $5.97\cdot 10^{24}$ \\
 $a$ & \dotfill semimajor axis & m & $3.84399\cdot 10^{8}$\\
 $n$ & \dotfill mean motion, i.e. $2\pi/P_{\rm orb}$ & yr$^{-1}$ & 84\\
 $e$ & \dotfill orbital eccentricity & & 0.0549\\
 $(B-A)/C$ & \dotfill triaxiality & & $2.278\cdot 10^{-4}$ \\
 ${G}$ & \dotfill gravitational constant & m$^3$ kg$^{-1}$ yr$^{-2}$ & $66468$\\
 $\tau_M$ & \dotfill Maxwell time (Ratio of viscosity to unrelaxed rigidity) & yr &  8\\
 $\mu$ & \dotfill unrelaxed rigidity modulus & Pa
  & $0.8\cdot10^{11}$\\
 $\alpha$ & \dotfill the Andrade parameter & & $0.2$\\
 \hline
 \label{table}
 \end{tabular}
 \end{table*}

\begin{figure*}
\begin{minipage}{82mm}
\includegraphics[width=82mm]{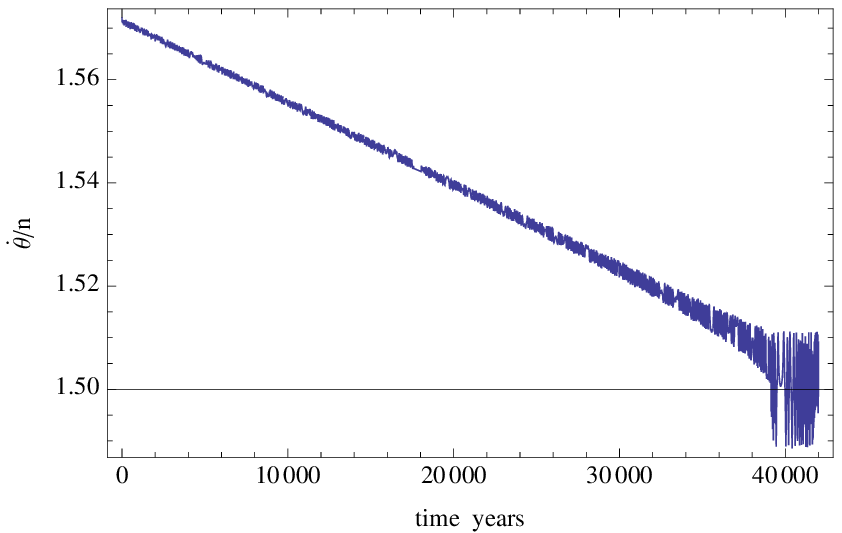}
\end{minipage}\hspace{2pc}
\begin{minipage}{82mm}
\includegraphics[width=82mm]{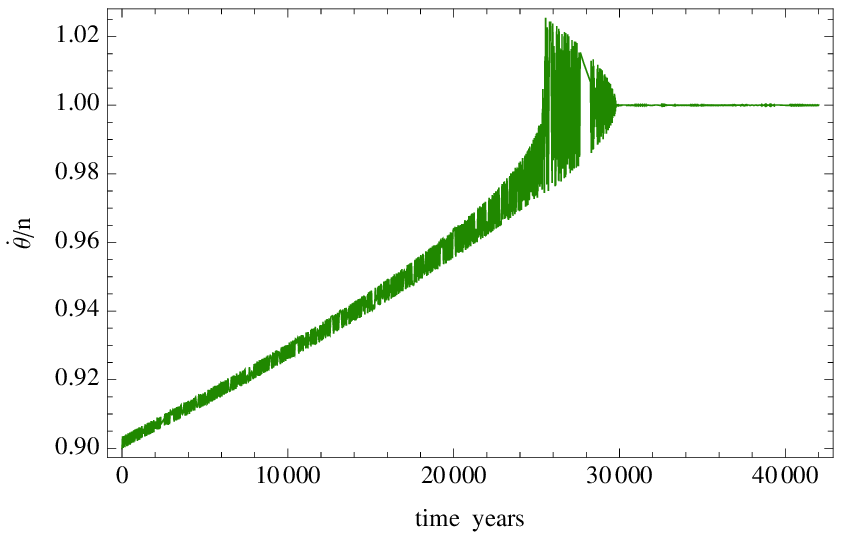}
\end{minipage}
\caption{Capture of the Moon  in numerical simulations into a) 3:2 spin-orbit resonance from faster 
prograde rotation; b) 1:1 resonance from initially slower prograde rotation.}
\label{capt.fig}
\end{figure*}
Fig. \ref{capt.fig}a shows a typical example of numerical integration of the Moon's spin rate, which
includes both secular and oscillating components of tides raised by the Earth, as well as the
traxiality-induced torques. The initial rate is $\dot\theta=1.572\,n$, and the maximum step of integration
is $1.5\cdot10^{-4}$ yr. The plot shows the characteristic features of resonance capture: the spin 
rate decelerates
steadily and at a nearly constant rate on this timescale, the amplitude of free librations grows
toward the resonance and suddenly doubles upon the capture, after which it starts to decline due to the
dissipation of kinetic energy. More remarkable is the fact that the Moon is captured into the 3:2 resonance,
despite its nearly circular orbit. Thus, capture of the Moon into supersynchronous resonances is possible
with the present-day parameters.

\section{Probabilities of capture}
There are two ways of estimating the probability of capture into a spin-orbit resonance with a given
set of parameters. The first way is brute-force integrations of the differential equation of second order
for the angular acceleration caused by the polar component of the tidal torque acting on the Moon, and
for a grid of initial phase space parameters $\{\theta,\dot\theta\}$.
For the sake of simplicity, but without a loss of generality, these  integrations are started at zero
mean anomaly, ${\cal M}$$ (0)=0$, i.e., at perigees. The implicit assumption used in this method is that
any sidereal azimuthal angle $\theta$ is equally likely for a given spin rate $\dot\theta$ when the Moon
passes through a perigee. I performed
small-scale simulations, integrating the corresponding second-order ODE twenty times for these initial
parameters: $\dot\theta(0)=1.572\,n$, ${\cal M}(0)=0$, $\theta(0)=(j-1)\pi/20$, $j=1,2,\ldots,20$, the Maxwell
time being fixed at 8 yr. I found 12 captures and 8 passages, resulting in a capture probability of roughly
$0.6$. 

\begin{figure*}
\begin{minipage}{82mm}
\includegraphics[width=82mm]{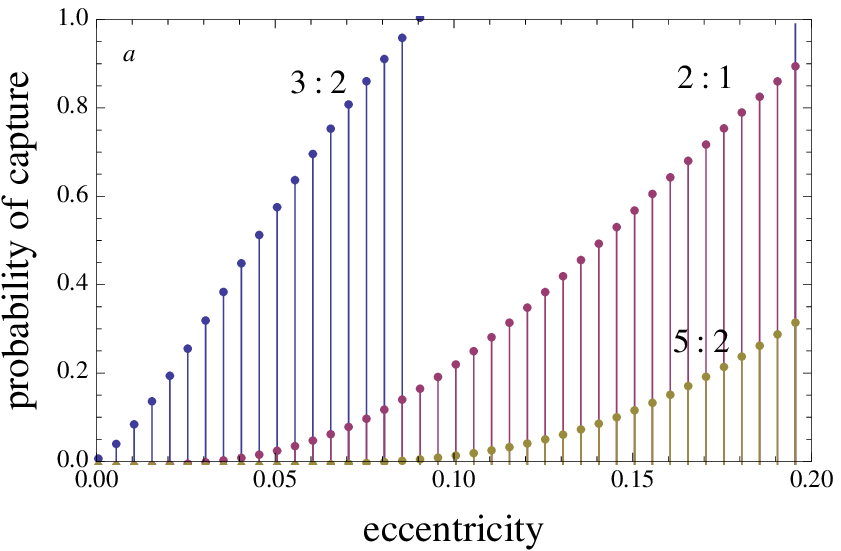}
\end{minipage}\hspace{2pc}
\begin{minipage}{82mm}
\includegraphics[width=82mm]{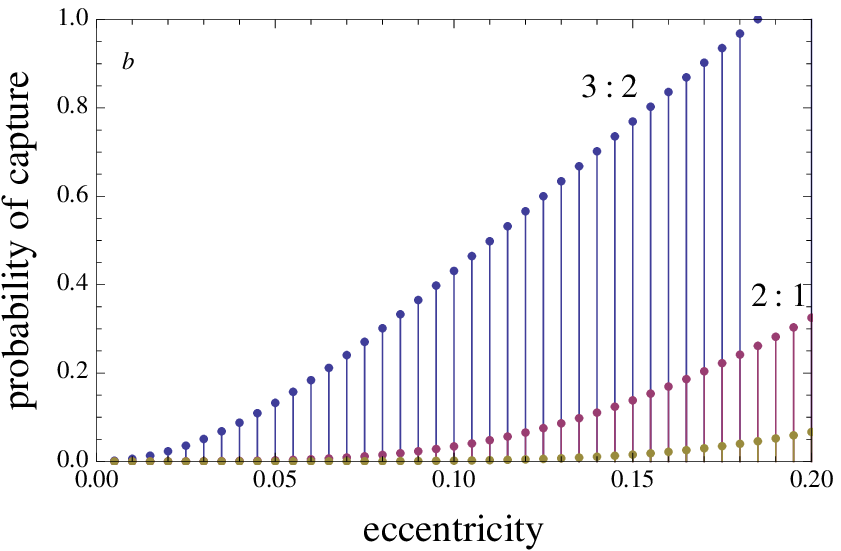}
\end{minipage}
\caption{Probabilities of capture of the Moon into the 3:2, 2:1, and 5:2 spin-orbit resonances
as functions of eccentricity, for a) $\tau_M=8$ yr and b) $\tau_M=500$ yr.}
\label{prob.fig}
\end{figure*}

The other way of estimating capture probabilities is the adaptation of the derivation proposed by
\citet{gol68} for the constant phase lag and the constant time lag tidal models. The details
of this calculation are given in \citep{ma12} and, in greater detail, in
\citep{mabe}. Fig. \ref{prob.fig}a and \ref{prob.fig}b depict
the results for two characteristic values of $\tau_M$, 8 yr and 500 yr, respectively. The results also
depend on the measure of quadrupole elongation, $(B-A)/C$, but to a lesser degree. It should be noted that
this semi-analytical calculation is based on the assumption that the energy 
offset from zero
at the beginning of the last libration above the resonance is uniformly distributed between
$0$ and the total energy dissipated by the secular tidal torque along the separatrix trajectory during one free
libration cycle \citep[see, e.g.,][]{pea}. 
This assumption is probably quite good as long as the magnitude of the permanent figure's torque
is much greater than the magnitude of tidal torques. Caution should be exercised with this approach for
nearly axially-symmetrical bodies, which are more easily captured into spin-orbit resonances, all other
parameters being the same. The strong nonlinearity of the tidal force may skew the probability distribution
of the residual rotational energy at the beginning of the last pre-resonance libration. Given this caveat,
we confirm that the capture probabilities strongly depend on the value of $\tau_M$. For example, as shown
in Fig.~\ref{prob.fig}, the probability of capture into 3:2 is $0.58$ for $\tau_M=8$ yr and $0.16$ for
$\tau_M=500$ yr.  At first glance, these numbers may seem to be consistent with the current state of
Moon's rotation, as the probability of traversing the higher resonances and entrapment in the 1:1 resonance
(which is always certain) is at least $\sim0.4$ for a wide range of the least-known parameter $\tau_M$. However,
recall that these estimates are obtained with the current low eccentricity. Why the high probabilities
of capture into a supersynchronous rotation represents a hard theoretical problem will be discussed in
\S\ref{discu.sec}. 

\section{Spin-down times}
\label{spindown.sec}
\begin{figure}
\includegraphics[width=82mm]{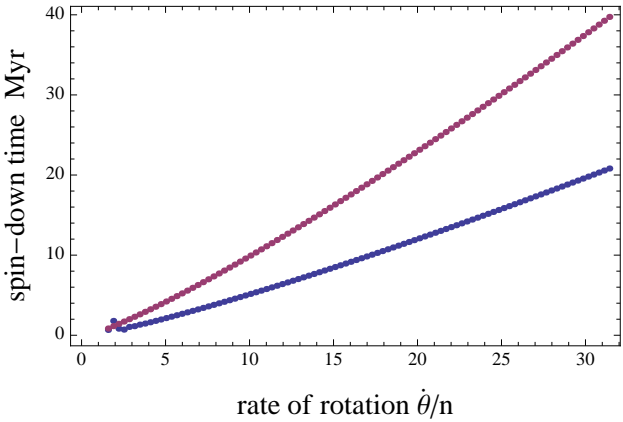}
\caption{Characteristic times of spin-down of the Moon for the current value of orbital eccentricity
$e=0.055$ (upper curve) and $e=0.3$ (lower curve). In both cases, the current value of semimajor axis
is assumed, and $\tau_M$ is set at 8 yr.}
\label{spindown.fig}
\end{figure}

The secular tidal torque in the model under investigation is negative at the present-day eccentricity
for any $\dot\theta>n$ except for
close vicinities of a few low-order spin-orbit resonances. Therefore, the general action of the tides
raised by the Earth on the Moon is to slow down the prograde rotation of the latter. Of special importance
is the characteristic spin-down time, which, following the previous literature \citep[e.g.,][]{ras}, 
can be defined as
\begin{equation}
t_\theta=\frac{\dot \theta}{|\ddot \theta_{\langle T \rangle}(\dot\theta)|}
\end{equation}
with $\ddot \theta_{\langle T \rangle}(\dot\theta)$ being the angular acceleration caused by the secular
tidal torque $\langle T \rangle$. In this computation, as ever, the obliquity of the lunar equator is
ignored. The results are shown in Fig. \ref{spindown.fig} for a grid of points in $\dot\theta/n$, chosen
in such a way as to avoid the sharp features at spin-orbit resonances, for two values of eccentricity:
$e=0.054$ and 0.3, and for the current value of semimajor axis, which is about $60$ Earth's radii.

Let us recall that in the "work-horse" tidal model of constant time lag (CTL), the deceleration of 
spin is arrested
when the state of pseudosynchronous rotation is reached at $\dot\theta_{\rm pseudo}/n\approx 1+6e^2$ \citep[e.g.,][]{hut}.
In reality, pseudosynchronous equilibria are unstable for terrestrial planets and moons \citep{me}.
Therefore, at small or moderate eccentricities, the Moon is bound to spin-down continuously until it is
captured into one of the spin-orbit resonances. Furthermore, most of the theories of Moon's origin suggest
that the Moon was formed much closer to the Earth than it is now \citep{can}. The characteristic spin-down times
are strong functions of the semimajor axis through the relation to the
polar torque, $\langle T \rangle \propto a^{-6}K_c(\dot\theta,n)$, where $K_c$ is the frequency-dependent
quality function defined in
\citep{efr2,ma12}. For example, if we compute the characteristic times for the same relative rates
$\dot\theta/n$ and $a=8R_{\rm Earth}$, we obtain practically the same curves as in Fig. \ref{spindown.fig},
but scaled down by approximately 5000. Observe that the dependence of tidal dissipation on $a$ is
much weaker here than in the CTL model, which predicts $t_\theta\propto a^6\dot\theta/n$ \citep{gol68,cor},
due to the fast decline of the quality function $K_c$ with tidal frequency. 

\section{Discussion}
One of the theoretical difficulties that the currently dominating giant impact theory of lunar formation
encounters is the excess angular momentum of the early Earth-Moon system. \citet{cuk} suggested a dynamical
scenario, which allows a fast-spinning proto-Earth to loose a sufficiently large amount of angular momentum
after a debris disk forming impact through a relatively short epoch of capture into the evection resonance.
As first suggested by Yoder in 1976, according to \citet{peca}, and mathematically developed by \citet{tou}, the lunar perigee is
locked in a synchronous precession with the orbital motion of the Earth around the Sun, and the long axis
of the lunar orbit stays at $90\degr$ from the Sun-Earth line. This resonance defeats the tidal actions
of circularization (secular decrease of eccentricity) and orbital expansion, allowing the eccentricity to
remain high for a span of time sufficiently long for the Earth to spin down. In the numerical simulation
presented by \citet{cuk}, the evection resonance holds for approximately 60 Kyr. Unfortunately, the
authors used a variant of the "constant-$Q$" model, which is not adequate for solid or partially melted
bodies \citep{em}. Their conclusions about the early dynamical evolution of the Moon-Earth system should be
taken with a grain of salt. The main difference between this {\it ad hoc} model and the realistic rheological
model is that in the latter, the quality function is a rising function at positive tidal frequencies asymptotically approaching zero (Fig. \ref{kink.fig}). The weakening of tidal dissipation at high tidal
frequencies may resolve the problem of overheating the Moon, as discussed in \S \ref{spindown.sec}. The
spin-down of the early Moon is still fast enough (a few Kyr) to justify the widely accepted assumption
in numerical simulations that the Moon is already synchronized by the time the evection resonance sets in. 
So much the more puzzling becomes the issue how the Moon traversed the higher spin-orbit resonances on
its way to synchronous rotation.

Indeed, capture into the 3:2 resonance becomes certain at $e\simeq0.09$ for $\tau_M=8$ yr, and $e=\simeq0.18$
for $\tau_M=500$ yr (Fig. \ref{prob.fig}). The simulations by \citet{cuk} suggest that the orbital eccentricity
acquires much higher values shortly after the onset of the evection resonance. Furthermore, these probabilities
are computed for the current mean motion of the Moon, whereas the giant impact theory implies much smaller
orbits for the early Moon, down to $4R_{\rm Earth}$. Somewhat counter-intuitively, the probabilities of capture
into a spin-orbit resonance become smaller for tighter orbits, all other parameters being the same. For
example, the probability of capture into 3:2 is only $0.2$ for the Moon at $a=4\,R_{\rm Earth}$, $\tau_M=8$ yr
and $e=0.055$. Could the Moon traverse the 3:2 resonance (and all the higher resonances) while it was
still very close to the Earth? Our calculations show that the Moon is inevitably entrapped in the 3:2
resonance at $a=4\,R_{\rm Earth}$, if the eccentricity exceeds $0.17$. But the evection mechanism quickly boosts
the orbital eccentricity to much higher values, up to $\simeq0.6$. Therefore, the only realistic possibility
for the Moon to avoid the 3:2 resonance within the giant impact scenario is to spin down to its present-day
synchronous state before the onset of the perigee precession resonance. This may take, depending on the initial
spin rate, up to 10 Kyr. This scenario also requires that the Moon remains fairly cold and viscous during
this pre-evection stage, which, due to the proximity to the Earth, may prove another hard problem. 

Simple calculations based on the formulae in \citet{peca} show that the dissipation of tidal energy in the Moon may exceed 
$10^{23}$ J yr$^{-1}$ for $a=10\,R_{\rm Earth}$, $\tau_M=8$ yr
and $e=0.055$ in the vicinity of the 3:2 resonance. This may raise the temperature of the Moon by $\simeq1$ 
K in 1 Kyr. The rise of temperature may be much faster at smaller distances from the Earth because of the
implicit $dE/dt\propto a^{-15/2}$ relation. For this calculation, I updated Eq. (31) in \citep{peca} by
including the realistic frequency-dependent quality function $K_c(\chi_{lmpq})$ instead of the constant quality
factor $\frac{3}{5}h_2/Q_{lmpq}$ used in that paper, and inserting the actual frequency mode. The latter update
takes into account that the original equation was derived specifically for the synchronous resonance. The resulting
general equation is:
\begin{eqnarray}
\langle \frac{dE}{dt}\rangle &=&  \vspace{-5mm}\frac{{\cal G}M_1^2R^5}{a^6}\; \sum_{m=0}^2 \frac{(2-m)!}{(2+m)!}(2-\delta_{0m}) \nonumber \\
&&\times\sum_{p=0}^2 \sum_{q=-\infty}^{+\infty}[F_{2mp}(i)\,G_{2pq}(e)]^2\,\chi_{2mpq}K_c(\chi_{2mpq})
\label{dedt.eq}
\end{eqnarray}
where ${\cal G}$ is the gravitational constant, $M_1$ is the mass of the Earth, $R$ is the radius of the Moon,
$a$ is the semimajor axis of the orbit, $i$ is the Moon's equator obliquity, $F_{2mp}(i)$ are the inclination functions,
$G_{2pq}(e)$ are the eccentricity functions, $\chi_{2mpq}=|\omega_{2mpq}|=|(2-2p+q)n-m\dot\theta|$ is the tidal frequency,
and $n$ is the orbital mean motion. This estimation is limited to the leading degree $l=2$, because the higher-degree terms
are smaller in amplitude by at least several orders of magnitude. The specific equations for the quality function can be found
in \citep{efr2}. One of the essential differences between Eq. \ref{dedt.eq} and Eq. (31) in \citep{peca} is the positively
defined tidal frequency $\chi_{2mpq}$ in the former replacing the factor $(2-2p+q-m)\,n$ in the latter, which can change sign.
An accurate derivation of the $dE/dt$ equation shows that the rate of tidal dissipation is proportional to $\omega_{lmpq}k_l(\omega_{lmpq})
\sin\epsilon_l(\omega_{lmpq})$, where $k_l$ is the frequency-dependent dynamical Love number, and $\epsilon_l$ is the degree-$l$
phase lag. Observing that $k_l$ is an even function of the tidal mode, and $\epsilon_l$ is an odd function, this product
can be more concisely written as the positively-defined function $\chi_{2mpq}K_c(\chi_{2mpq})$ of the physical frequency.
The resulting rate of tidal heating from Eq. \ref{dedt.eq} may therefore be significantly higher than the previously published estimates.
The leading terms of the quality function $K_c(\chi_{2mpq})$ vanish at the corresponding tidal modes, for example, $K_c(\chi_{2200})=0$
for $\dot\theta=1\,n$, turning to zero the tidal torque and acceleration. That does not, however, imply that tidal dissipation almost
ceases when the planet is locked in a spin-orbit resonance. The presence of other $lmpq$-modes, multiplied by their
tidal frequencies, makes up for a significant net dissipation. The character of the tidal heating versus spin rate dependence
is distinctly different with this model, to be discussed elsewhere.

\citet{peca} briefly mention the possibility that the Moon was locked into the 3:2 spin-orbit resonance for
a finite time span. A similar suggestion was made by \citet{garr}, who also found evidence for a high-eccentricity episode
in the dynamical history of the Moon from its present-day shape. Capture into a spin-orbit resonance should have happened before or 
at the very beginning of the evection
resonance, while the distance to the Earth remained small. If the subsequent rise of eccentricity
finds the Moon still in the 3:2 spin-orbit resonance, the tidal dissipation becomes a few orders of
magnitude stronger, and a complete or partial melt-down may follow. For example, the dissipation
for $a=10\,R_{\rm Earth}$, $\tau_M=8$ yr and $e=0.5$ is $\sim10^{24.5}$ J yr$^{-1}$. This would be sufficient
to heat the Moon by $3.6$ K per century. If the epoch of high eccentricity during the evection resonance lasts
for 40 Kyr, the temperature rises by $\sim1440$ K, which is above the melting point of silicates, including olivine and pyroxene.
Besides, it is not
obvious what kind of dynamical action could drive the Moon out of the resonance, apart from a fortuitous
high-velocity impact from an external body. Once captured into a spin-orbit resonance, a triaxial body can
traverse it only through a small opening in the phase space \citep{ma12}. In particular, the angle between
the "long" axis of the body and the center line should reach nearly $90\degr$ at perigee for this to happen.
Upon capture into the 3:2 resonance, the amplitude of the angle variation is close to that threshold value,
but the lunar free librations are damped quickly because of the high tidal dissipation, and the forced
librations are usually insignificant. Beyond the evection resonance, the eccentricity is bound to decrease,
further reducing the amplitude of forced librations.

Outside of the giant impact hypothesis of lunar origin, other plausible scenarios exist, which are consistent
with the current state of the satellite. If the Moon always remained on a low-eccentricity orbit during
the initial spin-down epoch, and it was cold and unyielding to the tidal forces, it could naturally
traverse the supersynchronous resonances before settling in the 1:1 resonance. Alternatively, the Moon could have a retrograde rotation at its formation. The tidal pull of the Earth in this case will slow down the
retrograde spin, and then will spin the Moon up in the prograde direction, until it falls into the 1:1
resonance, as shown in Fig. \ref{kink.fig}b. The only obstacle on this way is the subsynchronous 1:2
resonance. This resonance, however, is significantly weaker then the 1:1 and 3:2 resonances, and an unhindered
passage is secured with not too high eccentricities.
\label{discu.sec}

\label{lastpage}

\end{document}